\documentclass[notitlepage]{article}
\usepackage{amsmath}

\input{tcilatex}

\begin{document}

\title{Two Similarity Reductions and New Solutions for the Generalized
Variable-Coefficient KdV Equation by Using Symmetry Group Method}
\author{Rehab M. El-Shiekh$^{1,2}$ \\
$^{1}$Department of Mathematics, Faculty of Education,\\
Ain Shams University, Cairo, Egypt.\\
$^{2}$Department of Mathematics,College of Science and Humanities \\
at Howtat Sudair, Majmaah University, Kingdom of Saudi Arabia}
\date{}
\maketitle

\begin{abstract}
In this paper, a generalized variable-coefficient KdV equation (vcKdV)
arising in fluid mechanics, plasma physics and ocean dynamics is
investigated by using symmetry group analysis. Two basic generators are
determined, and for every generator in the optimal system the admissible
forms of the coefficients and the corresponding reduced ordinary
differential equation are obtained. The search for solutions to those
reduced ordinary differential equations yields new exact solutions for the
generalized vcKdV equation.
\end{abstract}


\section{Introduction}

\bigskip In 1895, Korteweg and de Vries developed a nonlinear partial
differential equation to model the propagation of shallow water waves [1].
This famous classical equation is known simply as the KdV equation.
Recently, the KdV equation have been derived and modified in many different
branches of science and engineering including the pulse-width modulation
[2], mass transports in a chemical response theory [3], dust acoustic
solitary structures in magnetized dusty plasmas [4] and nonlinear long
dynamo waves observed in the Sun [5]. However, the high-order nonlinear
terms must be taken into account in some complicated situations like at the
critical density or in the vicinity of the critical velocity [6--8]. The
modified KdV (mKdV)-typed equation, on the other hand, has recently been
used, e.g., to model the dust-ion-acoustic waves in such cosmic environments
as those in the supernova shells and Saturn's F-ring [9].

In many real physical backgrounds, the variable-coefficient nonlinear
evolution equations often can provide more powerful and realistic models
than their constant-coefficient counterparts when the inhomogeneities of
media and nonuniformities of boundaries are considered [10-22].Thereby, some
inhomogeneous KdV models with the time-dependent coefficients have been
derived to describe a variety of interesting and significant phenomena in
fluid mechanics, ocean dynamics and plasma physics, as follows:

In fluid mechanics, by considering the blood as incompressible fluid, the
evolution equation is obtained as a variable-coefficient mKdV (vcmKdV)
equation [14]%
\begin{equation}
U_{\tau }+\mu _{4}U^{2}U_{\zeta }+\mu _{2}U_{\zeta \zeta \zeta }+\mu
_{3}h_{2}(\tau )U_{\zeta }=0,
\end{equation}%
where $\mu _{2},\mu _{3}$ and $\mu _{4}$ are some constants describing the
material properties of the tube and the variable $U$ is the dynamical radial
displacement of the elastic tube.

In ocean dynamics, the pycnocline lies midway between the sea bed and
surface, the following vcmKdV model is obtained [8]%
\begin{equation}
u_{t}+a(t)u\text{ }u_{x}+k\text{ }u^{2}\text{ }u_{x}+u_{xxx}=0,
\end{equation}%
where $a(t)$ depends on the pycnocline location and k describes the internal
waves in a stratified ocean, as observed on the northwest shelf of Australia
[23] and in the Gotland deep of the Baltic Sea [24]. Also, in an
inhomogeneous two-layer shallow liquid, the governing equation modelling the
long wave propagation is the extended KdV model with variable coefficients
[25-26]%
\begin{equation}
u_{t}-6\alpha (t)u\text{ }u_{x}-6\gamma u^{2}\text{ }u_{x}+\theta
(t)u_{xxx}=0,
\end{equation}%
where $u(x,t)$ is proportional to the elevation of the interface between two
layers, $\alpha (t)$ and $\theta (t)$ imply that the ratio of the depths of
two layers depend on the coordinate \textquotedblleft t\textquotedblright\
[27].

In plasma physics, the investigation on the dynamics hidden in the plasma
sheath transition layer and inner sheath layer gave a perturbed mKdV model
[28],%
\begin{equation}
\psi _{\tau }+6\psi \text{ }\psi _{\eta }+\sigma \psi ^{2}\text{ }\psi
_{\eta }-\psi _{\eta \eta \eta }+h(\tau )\psi _{\eta }=0,
\end{equation}%
where $\sigma $ is a constant and $h(\tau )$ is an analytic function.

For the importance of the variable-coefficient KdV-type eqautions, we will
use the symmetry group analysis to study a generalized variable-coefficient
KdV equation (vcKdV) [29] 
\begin{equation}
u_{t}+g_{1}u_{xxx}+\left( g_{2}u^{3}+g_{3}u^{2}+g_{4}u+g_{5}\right)
u_{x}+g_{6}u+g_{7}=0,
\end{equation}%
where $g_{i}=g_{i}\left( t\right) ,$ $i=1,...,7,$ are arbitrary smooth
functions of $t$. Equation (5) includes considerably interesting equations
like equations (1-4). When, $g_{2}=0,$ equation $(5)$ is reduced to that in
Ref. $[30]$ derived by considering the time-dependent basic flow and
boundary conditions from the well-known Euler equation with an earth
rotation term. Also, if $g_{2}=g_{7}(t)=0,$ equation (5) becomes the
generalized Gardner equation with variable coefficients which arising in
nonlinear lattice, plasma physics and ocean dynamics [31]. Although the KdV
equation has been extensively studied for a long time via various methods,
yet the vcKdV equation has not been well handled as far as we known.
Recently, Qi Wang in $\left[ 29\right] $ used the semi-inverse method to
obtain the variational principle for it but no solutions obtained.
Therefore, the most important target for this paper is obtaining new exact
solutions for equation (5) under some constraints among the variable
coefficients by using the symmetry group analysis.

\section{Symmetry method}

We briefly outlined Steinberg's similarity method of finding explicit
solutions for both linear and non-linear partial differential equations $%
[32-37]$. The method based on finding the symmetries of the differential
equations is as follows:\newline
Suppose that the differential operator $L$ can be written in the form%
\begin{equation}
L(u)=\frac{\partial ^{p}u}{\partial t^{p}}-H(u),
\end{equation}%
\newline
\newline
where $u=u(t,x)$ and $H$ may depend on $t,x,u$ and any derivative of $u$ as
long the derivative of $u$ dose not contain more than $(p-1)$ $,t$
derivatives. We will consider the symmetry operator (called infinitesimal
symmetry), being quasi-linear partial differential operator of first-order,
in the form of\newline
\begin{equation}
S(u)=A(t,x,u)\frac{\partial u}{\partial t}+\tsum_{i=1}^{n}B_{i}(t,x,u)\frac{%
\partial u}{\partial x_{i}}+C(t,x,u),
\end{equation}%
\newline
\newline
and define the Fr\`{e}chet derivative of $L(u)$ by\newline
\begin{equation}
F(L,u,v)=\frac{d}{d\varepsilon }L(u+\varepsilon v)|_{\varepsilon =0}
\end{equation}%
\newline
\newline
With these definitions, we will conduct the following steps:\newline
$(i)$ Compute $F(L,u,v);\newline
(ii)$ Compute $F(L,u,S(u))$;\newline
$(iii)$ Substitute $H(u)$ for ($\frac{\partial ^{p}u}{\partial t^{p}})$ in $%
F(L,u,S(u));\newline
(iv)$ Set this expression to zero and perform a polynomial expansion;\newline
$(v)$ Solve the resulting partial differential equations. Once this system
of partial differential equations is solved for the coefficients of $S(u)$,
Eq. $(6)$ can be used to obtain the functional form of the solutions.

\section{Determination of the symmetries}

The vcKdV equation can be expressed as%
\begin{equation}
L(u)=u_{t}+g_{1}u_{xxx}+\left( g_{2}u^{3}+g_{3}u^{2}+g_{4}u+g_{5}\right)
u_{x}+g_{6}u+g_{7}=0.
\end{equation}

In order to find the symmetries of Eq. $\left( 9\right) $, we set the
following symmetry operator%
\begin{equation}
S\left( u\right) =A\left( x,t,u\right) u_{t}+B\left( x,t,u\right)
u_{x}+C\left( x,t,u\right) .
\end{equation}%
Calculating the Fr\'{e}chet derivative $F\left( L,u,v\right) $ of $L\left(
u\right) $ in the direction of $v,$ given by Eq. $\left( 8\right) $, and
replacing $v$ \ by $S\left( u\right) $ in $F,$ we get%
\begin{gather}
F\left( L,u,S\left( u\right) \right)
=S_{t}+g_{1}S_{xxx}+3g_{2}u^{2}u_{x}S+g_{2}u^{3}S_{x}+g_{3}u^{2}S_{x}  \notag
\\
+2g_{3}u\text{ }u_{x}S+g_{4}u_{x}S+g_{4}uS_{x}+g_{5}S_{x}+g_{6}S+g_{7}.
\end{gather}%
Substituting the values of different derivatives of $S(u)$ in $F$ with the
aid of Maple program, we get a polynomial expansion in $%
u_{x},u_{t},u_{y},u_{x}u_{t}$,...,etc. On making use of Eq. $(9)$ in the
polynomial expression for $F$, rearranging terms of various powers of
derivatives of $u$ and equating them to zero, we obtain%
\begin{gather}
A_{x}=A_{u}=B_{u}=C_{u\text{ }u}=C_{xu}=0,  \notag \\
3g_{1}B_{x}-\left( Ag_{1}\right) _{t}=0,  \notag \\
C_{t}-\left( Ag_{7}\right) _{t}-\left( Ag_{6}\right)
_{t}u+g_{1}C_{xxx}+g_{2}C_{x}u^{3}+g_{3}C_{x}u^{2}+g_{4}C_{x}u  \notag \\
+g_{5}C_{x}-g_{6}C_{u}u-g_{7}C_{u}+g_{6}C=0,  \notag \\
g_{2}B_{x}u^{3}+g_{5}B_{x}-\left( Ag_{5}\right) _{t}-\left( Ag_{2}\right)
_{t}u^{3}-\left( Ag_{3}\right) _{t}u^{2}+2g_{3}uC+B_{t}+g_{4}C  \notag \\
+3g_{2}u^{2}C-\left( Ag_{4}\right) _{t}u+g_{4}B_{x}u+g_{3}B_{x}u^{2}=0.
\end{gather}%
On solving system $\left( 12\right) ,$ the infinitesimal $A,B$ and $C$ in
the above equations are:%
\begin{eqnarray}
A &=&\frac{1}{\Gamma ^{\prime }\left( t\right) }[3c_{1}\Gamma \left(
t\right) +c_{2}],\text{ \ \ }\frac{d\Gamma \left( t\right) }{dt}=g_{1}\left(
t\right) ,  \notag \\
B &=&c_{1}x+c_{3},  \notag \\
C &=&c_{4}u+c_{5},
\end{eqnarray}%
where $c_{i},i=1,2,...,5,$ are arbitrary constants. The functions $%
g_{i}=g_{i}\left( t\right) ,i=1,2,...,7,$ are governed by the following
equations:%
\begin{eqnarray}
\left( Ag_{2}\right) _{t}-\left( c_{1}+3c_{4}\right) g_{2} &=&0,  \notag \\
\left( Ag_{3}\right) _{t}-\left( c_{1}+2c_{4}\right) g_{3}-3c_{5}g_{2} &=&0,
\notag \\
\left( Ag_{4}\right) _{t}-\left( c_{1}+c_{4}\right) g_{4}-2c_{5}g_{3} &=&0, 
\notag \\
\left( Ag_{5}\right) _{t}-c_{1}g_{5}+c_{5}g_{4} &=&0,  \notag \\
\left( Ag_{6}\right) _{t} &=&0,  \notag \\
\left( Ag_{7}\right) _{t}+c_{4}g_{7}+c_{5}g_{6} &=&0.
\end{eqnarray}%
The symmetry Lie algebra $\tciLaplace ^{5}$ of Eq. $(5)$ is generated by the
operators%
\begin{eqnarray}
V_{1} &=&\frac{3\Gamma \left( t\right) }{\Gamma ^{\prime }\left( t\right) }%
\frac{\partial }{\partial t}+x\frac{\partial }{\partial x},  \notag \\
V_{2} &=&\frac{1}{\Gamma ^{\prime }\left( t\right) }\frac{\partial }{%
\partial t},  \notag \\
V_{3} &=&\frac{\partial }{\partial x},  \notag \\
V_{4} &=&u\frac{\partial }{\partial u},  \notag \\
V_{5} &=&\frac{\partial }{\partial u}.
\end{eqnarray}%
where $\tciLaplace ^{5}$ is the direct sum of $V_{1},...,V_{5}$ and the
commutator table of it is given by

\begin{center}
\begin{tabular}{|p{0.65in}|p{0.65in}p{0.65in}p{0.65in}p{0.65in}p{0.65in}|}
\hline
& $V_{1}$ & $V_{2}$ & $V_{3}$ & $V_{4}$ & $V_{5}$ \\ \hline
$V_{1}$ & $0$ & $-3V_{2}$ & $-V_{3}$ & $0$ & $0$ \\ 
$V_{2}$ & $3V_{2}$ & $0$ & $0$ & $0$ & $0$ \\ 
$V_{3}$ & $V_{3}$ & $0$ & $0$ & $0$ & $0$ \\ 
$V_{4}$ & $0$ & $0$ & $0$ & $0$ & $-V_{5}$ \\ 
$V_{5}$ & $0$ & $0$ & $0$ & $V_{5}$ & $0$ \\ \hline
\end{tabular}
\end{center}

\section{One-dimensional optimal system of $\tciLaplace ^{5}$}

We need to search a one-dimensional optimal system of $\tciLaplace ^{5}$by
considering a general element of $\tciLaplace ^{5}$, $V=\sum%
\limits_{i=1}^{5}a_{i}V_{i}$, and checking whether $V$ can be mapped to a
new element $V^{\ast }$under the general adjoint transformation $Ad\left(
\exp \left( \varepsilon V_{i}\right) \right) V_{j}=V_{j}-\varepsilon \left[
V_{i},V_{j}\right] +\frac{\varepsilon ^{2}}{2}\left[ V_{i},\left[ V_{i},V_{j}%
\right] \right] ,$ so as to simplify it as much as possible.

\begin{center}
\ \ \ The adjoint table of $\tciLaplace ^{5}$

\begin{tabular}{|p{0.65in}|p{0.65in}p{0.65in}p{0.65in}p{0.65in}p{0.65in}|}
\hline
& $V_{1}$ & $V_{2}$ & $V_{3}$ & $V_{4}$ & $V_{5}$ \\ \hline
$V_{1}$ & $V_{1}$ & $\exp \left( 3\varepsilon \right) V_{2}$ & $\exp \left(
\varepsilon \right) V_{3}$ & $V_{4}$ & $V_{5}$ \\ 
$V_{2}$ & $V_{1}-3\varepsilon V_{2}$ & $V_{2}$ & $V_{3}$ & $V_{4}$ & $V_{5}$
\\ 
$V_{3}$ & $V_{1}-\varepsilon V_{3}$ & $V_{2}$ & $V_{3}$ & $V_{4}$ & $V_{5}$
\\ 
$V_{4}$ & $V_{1}$ & $V_{2}$ & $V_{3}$ & $V_{4}$ & $\exp \left( \varepsilon
\right) V_{5}$ \\ 
$V_{5}$ & $V_{1}$ & $V_{2}$ & $V_{3}$ & $V_{4}-\varepsilon V_{5}$ & $V_{5}$
\\ \hline
\end{tabular}%
\\[0pt]

\bigskip
\end{center}

Following $[38],$ we can deduce the following basic fields which form an
optimal system for the coupled KdV system,\newline
$(i)$ $V_{1}+k_{1}V_{4,}$\newline
$\left( ii\right) V_{2}+k_{2}V_{3}+k_{3}V_{4,}$\newline
$\left( iii\right) V_{3}+k_{4}V_{5},$\newline
$\left( iv\right) V_{4},$\newline
$\left( v\right) V_{5},$\newline
where $k_{i},i=1,...,4$ are arbitrary constants. Because cases $(iii)$,$(iv)$
and $(v)$ give trivial reductions therefore, we will discuss the first and
the second cases only.

In order to obtain the invariant transformation in each of the above cases,
we write the characteristic equation in the form%
\begin{equation}
\frac{dt}{A\left( x,t,u\right) }=\frac{dx}{B\left( x,t,u\right) }=\frac{-du}{%
C\left( x,t,u\right) }.
\end{equation}

\section{Reductions and exact solutions}

In this section, the primary focus is on the reductions associated with the
two vector fields (i) and (ii), and the admissible forms of the coefficients
by solving Eq. (16) and the system of integrability conditions (14)

\underline{\textbf{Generator (I)}}

The generator $(i)$ in the optimal system defines the similarity variable
and the similarity solution as follows:%
\begin{eqnarray}
\zeta &=&(x+k_{1})\Gamma \left( t\right) ^{\frac{-1}{3}},  \notag \\
u\left( x,t\right) &=&F\left( \zeta \right) \Gamma \left( t\right) ^{\frac{%
-k_{1}}{3}},
\end{eqnarray}%
and the coefficients are given as:%
\begin{eqnarray}
g_{2}\left( t\right) &=&\frac{n_{1}}{3}\Gamma ^{\prime }\left( t\right)
\Gamma \left( t\right) ^{\frac{-2}{3}+k_{1}},  \notag \\
g_{3}\left( t\right) &=&\frac{n_{2}}{3}\Gamma ^{\prime }\left( t\right)
\Gamma \left( t\right) ^{\frac{2}{3}(k_{1}-1)},  \notag \\
g_{4}\left( t\right) &=&\frac{n_{3}}{3}\Gamma ^{\prime }\left( t\right)
\Gamma \left( t\right) ^{\frac{1}{3}(k_{1}-2)},  \notag \\
g_{5}\left( t\right) &=&\frac{n_{4}}{3}\Gamma ^{\prime }\left( t\right)
\Gamma \left( t\right) ^{-\frac{2}{3}},  \notag \\
g_{6}\left( t\right) &=&\frac{n_{5}}{3}\Gamma ^{\prime }\left( t\right)
\Gamma \left( t\right) ^{-1},  \notag \\
g_{7}\left( t\right) &=&\frac{n_{6}}{3}\Gamma ^{\prime }\left( t\right)
\Gamma \left( t\right) ^{-(1+\frac{k_{1}}{3})},
\end{eqnarray}%
where $n_{1},...,n_{6}$ are arbitrary constants. Using the similarity
variable, the forms of the similarity solution and the coefficient
functions, the vcKdV is reduced to the following ordinary differential
equation.%
\begin{equation}
3F^{\prime \prime \prime }+n_{1}F^{3}F^{\prime }+n_{2}F^{2}F^{\prime
}+n_{3}FF^{\prime }+(n_{4}-\zeta )F^{\prime }+(n_{5}-k_{1})F+n_{6}=0.
\end{equation}%
To solve Eq. $(19)$, we seek a special solution in the form of%
\begin{equation}
F=A_{0}+A_{1}\zeta ^{-\frac{2}{3}},
\end{equation}%
where $A_{0}$ and $A_{1}$are arbitrary constants to be determined.
Substituting Eq. (20) into Eq. (19) and equating the coefficients of
different powers of $\zeta $ to zero, we get a system \ of algebraic
equations, solutions of which give rise to the relations on the constants as 
\begin{equation}
A_{0}=\frac{3n_{6}}{2},\text{ }A_{1}=\left( \frac{60n_{6}}{n_{2}}\right) ^{%
\frac{1}{3}},\text{ }n_{1}=-\frac{2n_{2}}{9n_{6}},\text{ }n_{3}=-\frac{3}{2}%
n_{2}n_{6},\text{ }n_{4}=\frac{3}{4}n_{2}n_{6}^{2},\text{ }k_{1}=\frac{2}{3}%
+n_{5}.
\end{equation}%
Finally, we get the following exact solution for the vcKdV%
\begin{equation}
u_{1}\left( x,t\right) =\Gamma \left( t\right) ^{\frac{-\left( \frac{2}{3}%
+n_{5}\right) }{3}}\left( \frac{3n_{6}}{2}+\left( \frac{60n_{6}}{n_{2}}%
\right) ^{\frac{1}{3}}\left( \left[ x+\frac{2}{3}+n_{5}\right] \Gamma \left(
t\right) ^{\frac{-1}{3}}\right) ^{-\frac{2}{3}}\right)
\end{equation}

\underline{\textbf{Generator (II)}}

Corresponding to this generator, the associated similarity variable and
similarity solution are given by%
\begin{eqnarray}
\zeta &=&k_{2}\Gamma \left( t\right) -x,  \notag \\
u\left( x,t\right) &=&F\left( \zeta \right) e^{^{-k_{3}\Gamma \left(
t\right) }},
\end{eqnarray}%
and the coefficient functions are given by%
\begin{eqnarray}
g_{2}\left( t\right) &=&m_{1}\Gamma ^{\prime }\left( t\right)
e^{^{3k_{3}\Gamma \left( t\right) }},  \notag \\
g_{3}\left( t\right) &=&m_{2}\Gamma ^{\prime }\left( t\right)
e^{^{2k_{3}\Gamma \left( t\right) }},  \notag \\
g_{4}\left( t\right) &=&m_{3}\Gamma ^{\prime }\left( t\right)
e^{^{k_{3}\Gamma \left( t\right) }},  \notag \\
g_{5}\left( t\right) &=&m_{4}\Gamma ^{\prime }\left( t\right) ,  \notag \\
g_{6}\left( t\right) &=&m_{5}\Gamma ^{\prime }\left( t\right) ,  \notag \\
g_{7}\left( t\right) &=&m_{6}\Gamma ^{\prime }\left( t\right)
e^{^{-k_{3}\Gamma \left( t\right) }},
\end{eqnarray}%
where $m_{1},...,m_{6}$ are arbitrary constants.

The reduced ordinary differential equation is%
\begin{equation}
F^{\prime \prime \prime }+m_{1}F^{3}F^{\prime }+m_{2}F^{2}F^{\prime
}+m_{3}FF^{\prime }+(m_{4}-k_{3})F^{\prime }+(k_{4}-m_{5})F+m_{6}=0.
\end{equation}%
Herein, we apply the modified extended tanh function method $[39-40]$ to
obtain exact solitary wave solutions to Eq. $(25)$ . Let us assume that Eq. $%
(25)$ has a solution in the form 
\begin{equation}
F\left( \zeta \right) =A_{0}+\sum_{i=1}^{N}A_{i}\phi ^{i}+B_{i}\phi ^{-i},
\end{equation}%
where $\phi \left( \zeta \right) $ is a solution of the following Riccati
equation%
\begin{equation}
\phi ^{\prime }=r+\phi ^{2},
\end{equation}%
which has the following solutions 
\begin{eqnarray}
\phi (\zeta ) &=&-\sqrt{-r}\tanh (\sqrt{-r}\zeta ),\text{ \ \ \ \ \ \ }r<0, 
\notag \\
\phi (\zeta ) &=&-\sqrt{-r}\coth (\sqrt{-r}\zeta ),\text{ \ \ \ \ \ \ }r<0, 
\notag \\
\phi (\zeta ) &=&\sqrt{r}\tan (\sqrt{r}\zeta ),\text{ \ \ \ \ \ \ \ \ \ \ \
\ \ \ \ }r>0,  \notag \\
\phi (\zeta ) &=&-\sqrt{r}\cot (\sqrt{r}\zeta ),\text{ \ \ \ \ \ \ \ \ \ \ \
\ }r>0,  \notag \\
\phi (\zeta ) &=&-\frac{1}{\zeta },\text{ \ \ \ \ \ \ \ \ \ \ \ \ \ \ \ \ \
\ \ \ \ \ \ \ \ \ \ \ }r=0.
\end{eqnarray}%
Substitute $\left( 26\right) $ into Eq. $\left( 25\right) $ and by balancing
the linear term with the greatest nonlinear term we get%
\begin{equation}
N=\frac{2}{3}
\end{equation}%
Therefore%
\begin{equation}
F\left( \zeta \right) =A_{0}+A_{1}\phi ^{\frac{2}{3}}+B_{1}\phi ^{\frac{-2}{3%
}}.
\end{equation}%
Substituting $\left( 30\right) $ into Eq. $\left( 25\right) $ and equating
the powers of $\phi ^{j},j=0,-\frac{11}{3},-3,-\frac{7}{3},...$ to zero, we
obtain a system of algebraic equations. By solving that system with Maple
program yields the following solution%
\begin{eqnarray}
A_{1} &=&-2\left( \frac{5}{9m_{1}}\right) ^{\frac{1}{3}},B_{1}=\left( \frac{%
15}{m_{1}}\right) ^{\frac{2}{3}}\left( \frac{3m_{3}m_{1}-m_{2}^{2}}{63m_{1}}%
\right) ,r=\pm \frac{1}{42}\sqrt{\frac{5}{14m_{1}}}\left( \frac{%
m_{2}^{2}-3m_{1}m_{3}}{m_{1}}\right) ^{\frac{3}{2}},  \notag \\
A_{0} &=&-\frac{m_{2}}{3m_{1}},m_{6}=0,k_{3}=m_{5},k_{2}=m_{4}+\frac{1}{%
1323m_{1}^{2}}\left( 98m_{2}^{3}-441m_{1}m_{2}m_{3}\pm 2\sqrt{70}\left(
m_{2}^{2}-3m_{1}m_{3}\right) ^{\frac{3}{2}}\right) .  \notag \\
&&
\end{eqnarray}%
Substituting Eq. $(31)$ into Eq. $(30),$ we get 
\begin{equation}
F\left( \zeta \right) =-\frac{m_{2}}{3m_{1}}-2\left( \frac{5}{9m_{1}}\right)
^{\frac{1}{3}}\phi ^{\frac{2}{3}}+\left( \frac{15}{m_{1}}\right) ^{\frac{2}{3%
}}\left( \frac{3m_{3}m_{1}-m_{2}^{2}}{63m_{1}}\right) \phi ^{\frac{-2}{3}},
\end{equation}%
where $\phi $ is given by Eq. $(28).$ Substitution of Eq. $(32)$ into Eq. $%
(23)$ results in%
\begin{gather}
u_{2}\left( x,t\right) =e^{^{-k_{3}\Gamma \left( t\right) }}\left[ \frac{%
-m_{2}}{3m_{1}}+2\left( \frac{5}{9m_{1}}\right) ^{\frac{1}{3}}r^{\frac{1}{3}%
}\tanh ^{\frac{2}{3}}(\sqrt{-r}\left( k_{2}\Gamma \left( t\right) -x\right)
)-\frac{\left( \frac{15}{m_{1}}\right) ^{\frac{2}{3}}\left( \frac{%
3m_{3}m_{1}-m_{2}^{2}}{63m_{1}}\right) }{r^{\frac{1}{3}}\tanh ^{\frac{2}{3}}(%
\sqrt{-r}\left( k_{2}\Gamma \left( t\right) -x\right) )}\right] ,  \notag \\
\\
u_{3}\left( x,t\right) =e^{^{-k_{3}\Gamma \left( t\right) }}\left[ \frac{%
-m_{2}}{3m_{1}}+2\left( \frac{5}{9m_{1}}\right) ^{\frac{1}{3}}r^{\frac{1}{3}%
}\coth ^{\frac{2}{3}}(\sqrt{-r}\left( k_{2}\Gamma \left( t\right) -x\right)
)+\frac{\left( \frac{15}{m_{1}}\right) ^{\frac{2}{3}}\left( \frac{%
3m_{3}m_{1}-m_{2}^{2}}{63m_{1}}\right) }{r^{\frac{1}{3}}\coth ^{\frac{2}{3}}(%
\sqrt{-r}\left( k_{2}\Gamma \left( t\right) -x\right) )}\right] ,  \notag \\
\end{gather}%
where%
\begin{equation*}
r=-\frac{1}{42}\sqrt{\frac{5}{14m_{1}}}\left( \frac{m_{2}^{2}-3m_{1}m_{3}}{%
m_{1}}\right) ^{\frac{3}{2}}\text{ and }k_{2}=m_{4}+\frac{1}{1323m_{1}^{2}}%
\left( 98m_{2}^{3}-441m_{1}m_{2}m_{3}-2\sqrt{70}\left(
m_{2}^{2}-3m_{1}m_{3}\right) ^{\frac{3}{2}}\right) ,
\end{equation*}%
\begin{eqnarray}
u_{4}\left( x,t\right) &=&-e^{^{-k_{3}\Gamma \left( t\right) }}\left[ \frac{%
m_{2}}{3m_{1}}+2\left( \frac{5}{9m_{1}}\right) ^{\frac{1}{3}}r^{\frac{1}{3}%
}\tan ^{\frac{2}{3}}(\sqrt{r}\left( k_{2}\Gamma \left( t\right) -x\right) )-%
\frac{\left( \frac{15}{m_{1}}\right) ^{\frac{2}{3}}\left( \frac{%
3m_{3}m_{1}-m_{2}^{2}}{63m_{1}}\right) }{r^{\frac{1}{3}}\tan ^{\frac{2}{3}}(%
\sqrt{r}\left( k_{2}\Gamma \left( t\right) -x\right) )}\right] ,  \notag \\
&& \\
u_{5}\left( x,t\right) &=&-e^{^{-k_{3}\Gamma \left( t\right) }}\left[ \frac{%
m_{2}}{3m_{1}}+2\left( \frac{5}{9m_{1}}\right) ^{\frac{1}{3}}r^{\frac{1}{3}%
}\cot ^{\frac{2}{3}}(\sqrt{r}\left( k_{2}\Gamma \left( t\right) -x\right) )-%
\frac{\left( \frac{15}{m_{1}}\right) ^{\frac{2}{3}}\left( \frac{%
3m_{3}m_{1}-m_{2}^{2}}{63m_{1}}\right) }{r^{\frac{1}{3}}\cot ^{\frac{2}{3}}(%
\sqrt{r}\left( k_{2}\Gamma \left( t\right) -x\right) )}\right] ,  \notag \\
&&
\end{eqnarray}%
where 
\begin{equation*}
r=\frac{1}{42}\sqrt{\frac{5}{14m_{1}}}\left( \frac{m_{2}^{2}-3m_{1}m_{3}}{%
m_{1}}\right) ^{\frac{3}{2}}\text{ and }k_{2}=m_{4}+\frac{1}{1323m_{1}^{2}}%
\left( 98m_{2}^{3}-441m_{1}m_{2}m_{3}+2\sqrt{70}\left(
m_{2}^{2}-3m_{1}m_{3}\right) ^{\frac{3}{2}}\right) ,
\end{equation*}%
\begin{equation}
u_{6}\left( x,t\right) =-e^{^{-k_{3}\Gamma \left( t\right) }}\left( \frac{%
m_{2}}{3m_{1}}+2\left( \frac{5}{9m_{1}}\right) ^{\frac{1}{3}}\left(
k_{2}\Gamma \left( t\right) -x\right) ^{\frac{-2}{3}}\right) ,
\end{equation}%
where 
\begin{equation*}
r=0,k_{2}=\frac{-m_{2}^{3}}{27m_{1}^{2}}+m_{4}\text{ and }m_{3}=\frac{%
m_{2}^{2}}{3m_{1}}.
\end{equation*}

\section{Conclusion}

In this paper, we have applied the symmetry group analysis to the vcKdV.
This application leads to two nonequavilant generatores, for every generator
in the optimal system the admissible forms of the coefficients and the
corresponding reduced ordinary differential equation are obtained. The
search for solutions to those reduced ordinary differential equations using
tanh function method has yielded many exact new solutions.

\end{document}